\documentclass[aps,12pt,nofootinbib]{revtex4}
\usepackage{graphics}
\usepackage[dvips]{color}
\usepackage{amsmath}
\usepackage{amssymb}

\usepackage{epsfig}

\def\beq{\begin{equation}}
\def\eeq{\end{equation}}

\def\bea{\arraycolsep .1em \begin{eqnarray}}
\def\eea{\end{eqnarray}}
\def\Tr{{\rm Tr}}
\newcommand{\step}{\vspace{.5em}}

\def\eps{\epsilon}
\def\om{\omega}

\def\eq#1{(\ref{#1})}

\def\s0#1#2{\mbox{\small{$ \frac{#1}{#2} $}}}
\def\0#1#2{\frac{#1}{#2}}

\def\grgl{\:\hbox to -0.2pt{\lower2.5pt\hbox{$\sim$}\hss}{\raise3pt\hbox{$>$}}\:}
\def\klgl{\:\hbox to -0.2pt{\lower2.5pt\hbox{$\sim$}\hss}{\raise3pt\hbox{$<$}}\:}

\newcommand \be {\begin{equation}}
\newcommand \ee {\end{equation}}
\newcommand \bed {\begin{displaymath}}
\newcommand \eed {\end{displaymath}}

\newcommand{\bit}{\begin{itemize}}
\newcommand{\eit}{\end{itemize}}


\begin{document}

\title{Ising exponents from the functional renormalisation group}

\author{Daniel F.~Litim${}^{a}$ and Dario Zappal\`a${}^{b}$}

\affiliation{
\mbox{${}^a$  Department of Physics and Astronomy, University of Sussex, BN1 9QH, Brighton, UK.}\\
\mbox{${}^b$ INFN, Sezione di Catania, 64 via S. Sofia, I-95123,
Catania, Italy.}
}%

\begin{abstract}
${}$\\[-1ex]

\centerline{\bf Abstract} We study the 3d Ising universality class
using the functional renormalisation group. With the help of
background fields and a derivative expansion up to fourth order we
compute the leading index, the subleading symmetric and anti-symmetric
corrections to scaling, the anomalous dimension, the scaling solution,
and the eigenperturbations at criticality.  We also study the
cross-correlations of scaling exponents, and their dependence on
dimensionality. We find a very good numerical convergence of the
derivative expansion, also in comparison with earlier findings.
Evaluating the data from all functional renormalisation group studies
to date, we estimate the systematic error which is found to be small
and in good agreement with findings from Monte Carlo simulations,
$\epsilon$-expansion techniques, and resummed perturbation theory.
\end{abstract}

\pagestyle{plain} \setcounter{page}{1}

\maketitle

\noindent{\bf 1. Introduction}\step

Continuous phase transitions of numerous systems in
statistical and particle physics belong to the
Ising universality class, characterised by the short range nature of
the interaction, a scalar order parameter, and the dimension.  The
absence of a physical length scale at the phase transition implies
scale invariance. Many fluids, magnets, or particle physics models
thus share the same behaviour at criticality described by universal
numbers such as the scaling exponent for the correlation length $\nu$,
its subleading correction $\omega$, and the anomalous dimension of the
order parameter at criticality $\eta$.  Further critical exponents for eg.~the 
specific heat $\alpha$, the spontaneous magnetisation $\beta$,
the magnetic susceptibility $\gamma$, and the magnetization at
criticality as a function of the external field $\delta$, are linked
to $\nu$ and $\eta$ by scaling relations \cite{ZinnJustin}.\step

The computation of universal indices -- in a quantum field theoretical
or statistical physics setting -- has become a benchmark test for
perturbative and non-perturbative methods in field theory. Accurate
predictions for exponents, scaling functions or amplitude ratios are
available based the renormalisation group, resummations of
perturbation theory, and numerical simulations on the lattice (see
\cite{Pelissetto:2000ek} for an overview).  An important continuum
method in the above is the functional renormalisation group
\cite{Polchinski:1983gv,Wetterich:1992yh}, based on the infinitesimal
integrating-out of momentum modes from a path integral representation
of the theory with the help of a Wilsonian momentum cutoff
\cite{Wilson:1973jj}.  By construction, functional flow equations
continuously interpolate between the microscopic theory at short
distances and the full quantum effective theory at large
distances. Powerful optimisation techniques are available to maximize
the physics content in given approximation, and to minimize cutoff
artifacts along the flow
\cite{Litim:2000ci,Litim:2001up,Litim:2001fd,Litim:2002cf,Pawlowski:2005xe}. A
particular strength of the functional renormalisation group is its
flexibility, allowing for perturbative as well as non-perturbative
approximations even in the presence of strong correlations or
couplings
\cite{Bagnuls:2000ae,Berges:2000ew,Polonyi:2001se,Pawlowski:2005xe,Litim:1998nf,Litim:2005us}.
\step

Fixed point studies for Ising-like theories have been performed within
various realisations of the functional renormalisation group including
Polchinski's formulation \cite{Bagnuls:2000ae}, Wetterich's equation
\cite{Berges:2000ew}, exact background field flows
\cite{Litim:2001ky,Litim:2002xm,Litim:2002hj,Litim:2006nn}, the
proper-time approximation
\cite{Liao:1996fp,Bohr:2000gp,Bonanno:2000yp,Mazza:2001bp,Litim:2001hk},
and discretized (hierarchical) transformations
\cite{Meurice:2007zg,Litim:2007jb}. The derivative expansion
\cite{Morris:1994ie,Litim:2001dt} and variations thereof
\cite{Berges:2000ew,Blaizot:2005xy} are the expansion schemes of
choice based on a small anomalous dimension. Scaling behaviour in more
complicated theories, eg.~thermal field theory \cite{Litim:1998yn},
gauge theories \cite{Bergerhoff:1995zq,Gies:2002af,Pawlowski:2003hq}
and gravity \cite{Litim:2008tt}, can equally be accessed using thermal
or gauge-covariant derivative and vertex expansions.\step

In this paper, we study the Wilson-Fisher fixed point in three
dimensions within a background field formulation.  In the past,
background field methods have mostly been employed for gauge theories
and gravity, where they allow for a gauge invariant implementation of
the cutoff \cite{Reuter:1993kw,Litim:1998nf,Freire:2000bq}, also
offering new expansions schemes \cite{Litim:2002ce}.  The motivation
for using this technique for non-gauge systems is two-fold. First of
all, the presence of a background field allows for a re-organisation
of the flow equation.  While this is of no relevance for the full
flow, it does make a difference once approximations are invoked. In
particular, derivative expansions of standard and background field
flows are different. This allows for complementary measurements of
universal scaling exponents.  Secondly, background field flows have
provided very good numerical results to lower orders in the derivative
expansion. Therefore it is important to understand whether this
pattern carries over to higher order.\step

In addition, we discuss the convergence of the derivative expansion
and provide an estimate for systematic uncertainties. Error estimates
are obtained by probing the dependence on the shape of the Wilsonian
cutoff function \cite{Litim:1996nw,Freire:2000sx} -- which vanishes
for the physical theory and hence should become small with increasing
order in the expansion -- and by checking the numerical convergence of
successive orders, which we extend up to fourth order.  We estimate
the systematic error by comparing different projections of the
Wilson-Fisher fixed point onto the flow equation, using a weighted
average over the available data.  We find a coherent picture and very
good agreement with the mean values and error estimates from Monte
Carlo and perturbative studies. We also evaluate the cross-dependences
of scaling exponents, and find an interesting link between the
expected error in an observable and its sensitivity to tiny variations
of the dimensionality. \step

The format of the paper is as follows. We recall the basic set-up
(Sec.~2) and our approximations (Sec.~3), followed by a discussion of
results (Sec.~4) and their optimisation (Sec.~5). Two sections deal
with the dependence on dimensionality (Sec.~6) and the
cross-correlation of exponents (Sec.~7). We evaluate the convergence
of the derivative expansion (Sec.~8) as well as systematic
uncertainties (Sec.~9), and close with a brief discussion
(Sec.~10).\\

\noindent{\bf 2. Renormalisation group}\step

Wilson's renormalisation group is based on the integrating-out of
momentum degrees of freedom from a path integral representation of the
theory. Modern, functional, implementations of this idea employ an
infrared momentum cutoff term $\Delta S_k=\s012\int _q\,
\varphi(q)\,R_k(q)\,\varphi(-q)$ for the propagating modes
$\varphi(q)$, added to the Schwinger functional with classical action
$S$ and external current $J$, 
\begin{equation}\label{Z} 
\ln Z_k[J]=\ln \int [d\varphi] 
\exp\left(-S-\Delta S_k +\int \varphi\cdot J\right)\,. 
\end{equation} 
The cutoff function $R_k(q)$ can be viewed as a momentum-dependent
mass term with $k$ denoting the RG momentum scale. It obeys
$R_k(q^2)\to 0$ for $k^2/q^2\to 0$ to ensure that the large momentum
modes $q^2\grgl k^2$ can propagate freely, and $R_k(q^2)>0$ for
$q^2/k^2\to 0$ which ensures that the low momentum modes $q^2\klgl
k^2$ are suppressed in the functional integral. This makes $R_k$ an
infrared cutoff. The change of \eq{Z} with the RG scale $k$ ($t=\ln
k$) reads $\partial_t Z_k=-\langle \partial_t \Delta S_k\rangle_J$. In
terms of the effective action $\Gamma_k[\phi]={\rm sup}_J(-\ln Z_k[J]
+\phi\cdot J)+\Delta S_k$ it is given by Wetterich's flow equation
\cite{Wetterich:1992yh} \be\label{flow} \partial_t\Gamma_k[\phi]=\s012
\Tr\frac{1}{\Gamma^{(2)}_k[\phi]+R_k}\partial_t R_k\,, \ee an exact,
functional differential equation which links the scale-dependence of
$\Gamma_k[\phi]$ with its second functional derivative
$\Gamma_k^{(2)}[\phi]\equiv\0{\delta^2\Gamma_k
  [\phi]}{\delta\phi\delta\phi}$ and (the scale-dependence of) the
regulator function $R_k$. The trace denotes a momentum integration,
and $\phi=\langle \varphi\rangle_J$ denotes the expectation value of
the field $\varphi$ at fixed external current $J$. By construction,
the flow \eq{flow} interpolates between an initial microscopic action
$\Gamma_\Lambda\approx S$ at $k=\Lambda$ and the full quantum
effective action at $k=0$. \step

Next, we discuss background field flows following
\cite{Litim:2001ky,Litim:2002xm,Litim:2002hj,Litim:2006nn} where a
non-propagating background field $\bar\phi$ is introduced into the
effective action $\Gamma_k[\phi]\to\Gamma_k[\phi,\bar\phi]$ by
coupling the fluctuation field $(\phi-\bar\phi)$ to the regulator and
the external current.  For the derivation of the flow, the background
field acts as a spectator, and we obtain \eq{flow} with the
replacement
$\Gamma^{(2)}_k[\phi]\to
\frac{\delta^2\Gamma_k[\phi,\bar\phi]}{\delta\phi\delta\phi}$.
The background field dependence of $\Gamma_k[\phi,\bar\phi]$ is
governed by
\begin{equation}
\frac{\delta}{\delta\bar\phi}\Gamma_k[\phi,\bar\phi]=
\frac{1}{2}\Tr\frac{1}{\Gamma^{(2)}_k[\phi,\bar\phi]+R_k}
\frac{\delta R_k[\bar\phi]}{\delta\bar\phi}
\end{equation}
and vanishes in the infrared limit $k\to 0$ where $R \to
0$. Subsequently, the background field will be identified with the
physical mean $\bar\phi=\phi$, leading to a background field flow for
an effective action $\Gamma_k[\phi]\equiv\Gamma_k[\phi,\phi]$. This
technique is standard practice in the study of gauge theories and
gravity leading to gauge-invariant flows within the background field
method.  In gauge and non-gauge theories, this procedure can simplify
the evaluation of the operator trace in \eq{flow}, which makes it
attractive for our purposes \cite{Litim:2001hk,Litim:2002hj}.  The key
difference between standard and background fields stems from the fact
that the presence of the background field, at an intermediate stage of
the computation, corresponds to a re-organisation the flow.  This
aspect is exploited below.\step

To be specific, we introduce the background field by substituting
$q^2\to\Gamma^{(2)}_k[\bar\phi,\bar\phi](q^2)$ in the regulator
function $R_k(q^2)$ (other choices such as $R_k\to R_k[\bar\phi]$ can
be used as well). For some class of $R_k$-functions
\cite{Litim:2002hj}, the flow \eq{flow} takes a very convenient form,
\begin{equation}\label{PT}
\partial_t\Gamma_k=
\Tr\left(\frac{k^2}{\Gamma^{(2)}_k/m+k^2}\right)^m +{\cal
O}(\partial_t\Gamma^{(2)}_k)\,. 
\end{equation} 
Here, $m\in[1,\infty]$ parametrises a remaining freedom in the choice
for the cutoff function which we fix later. The term $\propto
\partial_t\Gamma^{(2)}_k$ originates from the implicit $k$-dependence
introduced in $R_k$ via $\Gamma^{(2)}_k$, and reflects the
re-organisation of the flow through background fields.  The term
$\partial_t\Gamma^{(2)}_k$ on the r.h.s.~of \eq{PT} can be replaced
through a series -- starting off with the leading term in \eq{PT} and
functional derivatives thereof -- by making repeated use of \eq{PT}.
Closed forms for $\partial_t\Gamma_k$, or the flow for a few relevant
couplings are available under certain approximations
\cite{Litim:2002cf,Litim:2002hj,Gies:2002af}.  Below, we need the flow
for several field-dependent functions and therefore limit ourselves to
the leading term \cite{Litim:2002xm,Litim:2002hj}.\step
 
The first term on the r.h.s.~of \eq{PT}, or linear combinations for
various $m$, is equivalent to Liao's proper time flow equation
\cite{Liao:1996fp,Litim:2002xm,Litim:2006nn}
\begin{equation} \label{eq:gamma}
\partial_t \Gamma_k =
-\s012 {\rm Tr} \; \int_0^\infty \;\frac{ds}{s} \; (\partial_t f_k)\;
{\rm exp}\; \Big (-s
\frac{\delta^2 \Gamma_k}{\delta \phi\delta \phi}
\Big ) \,, 
\end{equation} 
originally derived from a proper time regularization of the one-loop
effective action. Eqs.~\eq{PT} and \eq{eq:gamma} are linked via
$f_k\equiv f_{\rm PT}(s\Lambda^2)-f_{\rm PT}(s k^2)$ with $f_{\rm
  PT}(x)=\Gamma(m,x)/\Gamma(m)$. The
flow \eq{PT} can equally be obtained from generalized Callan-Symanzik
flows \cite {Litim:2002xm} without the necessity for background
fields.  The flow equation \eq{PT} in the approximation \eq{eq:gamma}
has previously been used for studies of phase transitions
\cite{Bohr:2000gp,Bonanno:2000yp,Mazza:2001bp,Litim:2001hk,Litim:2001ky},
tunneling phenomena \cite{Zappala:2001nv}, spontaneous and
chiral symmetry breaking \cite{Bonanno:2004pq,Consoli:2006ji,Castorina:2003kq},
gravity \cite{Bonanno:2004sy}, and a general proof of convexity
\cite{Litim:2006nn}. Here, we will use it to analyse the infra-red
scaling at the Wilson-Fisher fixed point to fourth order in the
derivative expansion.\\

\noindent{\bf 3. Approximations}\step

In this section, we detail our ansatz for the effective action for a
real scalar field based on the derivative expansion and the relevant
renormalisation group equations.  With up to fourth order derivative
operators, the effective action reads 
\bea
\label{eq:formgam} \Gamma_k&=& \int d^Dx \; \Big \lbrack  V_k(\phi)+
\s012 Z_k(\phi) \,
\partial_\mu \phi\partial_\mu\phi+\; W_k (\phi) \; (\partial^2
\phi)^2 \Big \rbrack\,. 
\eea
The ansatz \eq{eq:formgam} should capture the relevant infrared
physics provided the anomalous dimension of the fields stay
small. Note that the derivative expansion has no small parameter
directly associated with it, because the integrand of \eq{flow}
receives dominant contributions for $q^2/k^2\klgl 1$
\cite{Litim:2001dt}.  Hence the numerical convergence has to be
checked a posteriori.  Good numerical convergence is known for
appropriate momentum cutoffs \cite{Litim:2000ci}. \step

The three functions $V, Z$ and $W$ in our ansatz \eq{eq:formgam} are
symmetric under reflection in field space $\phi\leftrightarrow
-\phi$. In principle, there are three independent tensor structures
available to fourth order in the derivative expansion,
\bea \label{eq:neglect} W_k (\phi) \;
(\partial^2 \phi)^2\,,\quad H_k(\phi)  \; \partial_\mu \phi
\,\partial_\mu\phi\, (\partial^2 \phi) \,,\quad\; J_k(\phi) \;
(\partial_\mu \phi\, \partial_\mu\phi)^2 \eea 
with $J$ $(H)$ symmetric (anti-symmetric) under reflection in field
space. In the free theory limit, the operators \eq{eq:neglect} scale
identically.  At an interacting fixed point this degeneracy is lifted,
and the higher derivative operators contribute with different
strengths to the flow \eq{PT}. We expect that the term $\sim W$ is the
most relevant one, for reasons detailed in Sect.~4. Therefore we
neglect $H$ and $J$. Then the initial conditions for the flow at
momentum scale $k=\Lambda$ are 
\beq\label{initial}
V_\Lambda(\phi)=\s012m^2_\Lambda\,\phi^2
+\s014\lambda_\Lambda\,\phi^4\,,\quad
Z_\Lambda(\phi)=1\,,\quad W_\Lambda(\phi)=0\,. 
\eeq 
For $k<\Lambda$, higher-order couplings are switched on due to the
renormalisation group running \eq{PT}, and the functions $V$, $Z$ and
$W$ develop a non-trivial field dependence. The Wilson-Fisher scaling
solution for $k\to 0$ corresponds to critical initial conditions
$m^2_{\Lambda,c}$ and $\lambda_{\Lambda,c}$. The renormalisation group
equations for the functions $V$, $Z$ and $W$ are obtained by inserting
\eq{eq:formgam} into \eq{eq:gamma} and expanding the exponential by
making use of the Baker-Campbell-Hausdorff formula.  The partial
differential equations for $V$, $Z$ and $W$ are of the form
\be\label{PDE} 
\partial_t X =-\frac{1}{2} \int_0^\infty \frac{ds}{s}
\; \int \frac{d^D p}{(2\pi)^D} \; (\partial_t f_k) \; e^{-s A_0}\,K_X
\ee 
where $X=V, Z$ or $W$, and $A_0=V''+Z\,p^2+2\,W\,p^4$, with primes on
functions denoting derivatives w.r.t.~the fields. The equations
\eq{PDE} encode the central physics of our setup.  The kernels $K_X$
encode the interactions amongst the operators in the ansatz
\eq{eq:formgam} under the renormalisation group. We have $K_V=1$. The
kernels $K_Z$ ($K_W$) are polynomials in the loop momentum variable
$p$ up to order $p^{14}$ ($p^{20}$) with coefficient functions
depending polynomially on $V, Z, W$ and their derivatives, and the
proper-time integration parameter $s$.  The expressions are very long
and not given explicitly.\step

For a fixed point study, it is convenient to introduce dimensionless
variables and a specific cutoff. We mainly use the parameter $1/m=0$
which is equivalent to the step function $f_{\rm PT}(y)=\theta\left
(1-y\right )$ with $y=s\, Z\, k^2$, achieved as $f_{\rm
  PT}(y)=\lim_{m\to\infty}\Gamma(m,m\,y)/\Gamma(m)$
\cite{Litim:2001ky} (see also \cite{Zappala:2002nx}). 
For large $\Gamma^{(2)}_k\gg k^2$, the flow then
becomes exponentially suppressed $\propto \exp(-\Gamma^{(2)}_k/k^2)$,
rather than algebraically. In consequence, amplitude expansions
convergence more rapidly \cite{Litim:2001dt}. The remaining
$s$-integration in \eq{PDE} is performed analytically. We introduce
\bea t&=&{\rm ln}(k/\Lambda)\,,\quad
x=k^{1-D/2-\eta/2}\,\phi\,,\quad\widehat p= p/k\,,\quad a_0=
A_0\,k^{-2+\eta}
\nonumber \\
v(x)&=&k^{-D}\, V(\phi)\,,\quad z(x)=k^{\eta}\, Z(\phi)\,,\quad
w(x)=k^{2+\eta}\, W(\phi)\,, \label{dimless} 
\eea 
where we have rescaled dimensionful variables in units of $k$;
$\eta=-\partial_t\ln Z_k$ denotes the anomalous dimension of the
field. It is understood that $v, z$ and $w$ are functions of $t$ and
$x$. Below, we denote derivatives w.r.t.~$x$ as eg.~$\partial_x
v\equiv v'$. In the parametrisation \eq{dimless}, the explicit
$k$-dependence of the differential equations \eq{PDE} is factored into
the variables.  We finally obtain 
\bea \partial_t Y+ D_Y\,Y - D_x ~x\,
Y'&=& \int \frac{d^D \widehat p}{(2\pi)^D} \; e^{- (a_0/z)}\,K_{Y} \,\
\label{zadim} \quad {\rm with}\quad Y=\{v',z,w\} 
\eea 
The terms on the left-hand sides display the canonical and anomalous
scaling of the fields $D_x=[\phi]$ and the variables $D_Y$, with
\beq
D_x=\s0{1}{2}(D-2+\eta)\,,\ \ D_{v'}=\s0{1}{2} (D+2-\eta)\,,\ \
D_z=- \eta \,,\ \ D_w=- (2+ \eta) \,. 
\eeq 
We note that the scaling dimensions $D_x$ and $D_{v'}$ ($D_z$ and
$D_w$) are positive (negative) for $\eta\ge 0$ and $D\ge2$. The terms
on the right-hand sides parametrise the non-trivial interactions
induced by \eq{eq:formgam} under the renormalisation group. The
integral kernels $K_Y$ are related to the kernels $K_X|_{s=1}$ in
\eq{PDE} via the relations $K_{v'}= -\partial_x \left ( {a_0}/{z}
\right )$, $K_z=K_Z\,k^{D+\eta}$, and $K_w=K_W\,k^{D+2+\eta}$.  \\

\begin{table}
\begin{center}
\begin{tabular}{cccccc}
 \hline
  \hline
  approximation&
${}\quad x_0\quad$ & $\quad v''(0)\quad$ & $\quad v''(x_0)\quad$ &
$\quad z(x_0)\quad$ & $\quad w(0)\quad$
\\ \hline
${}\quad$
LPA ${}\quad$ & $ 1.899$
 &
$-0.297$ & $0.672$ & $1$ & 0
\\
${}\quad$ ${\cal O}(\partial^2)$
${}\quad$ & $ 1.889$
 &
$-0.266$ & $ 0.601$ & $ 1.047$ & 0
\\
${}\quad$ ${\cal O}(\partial^4)$
${}\quad$ & $ 1.888$ & $ -0.267$ & $ 0.609$ & $ 1.050 $ & $-1.26\,
10^{-4}$
\\
 \hline \hline
 \end{tabular}
\end{center}
${}$\vskip-.5cm \caption{\label{results1} Potential minimum $x_0$,
curvature, and other reference values of the scaling solution.}
\end{table}

\noindent{\bf 4. Results}\step

In this section, we analyse the physics of \eq{zadim} at the
Wilson-Fisher fixed point, which, in $D=3$ dimensions, corresponds to
the unique non-trivial solution $Y_*(x)\neq 0$ of
\be\label{FP}
\partial_t Y _*= 0\,.
\ee 
In the limit of large $x\gg 1$, the r.h.s.~of \eq{zadim} is
exponentially suppressed and the fixed point solution is dominated by
the scaling of the fields and variables, \beq\label{scaling}
Y_*(x)\propto x^{D_Y/D_x}\,. \eeq Consequently, the solutions $z_*(x)$
and $w_*(x)$ vanish asymptotically because $D_z$ and $D_w$ are
$<0$. The algebraic suppression is the more pronounced the larger
$-D_Y$. For $v'$, we find a rising behaviour for large $x$ because
$D_{v'}>0$. For small $x\klgl 1$, the interaction terms become
relevant, and the complexity of the equations makes it necessary to
use numerical methods. Here, we solve \eq{zadim} with \eq{FP} for
$v_*'(x)$, $z_*(x)$ and $w_*(x)$ without making any further expansions
such as eg.~polynomial expansions. \step

\begin{figure}[t]
\epsfig{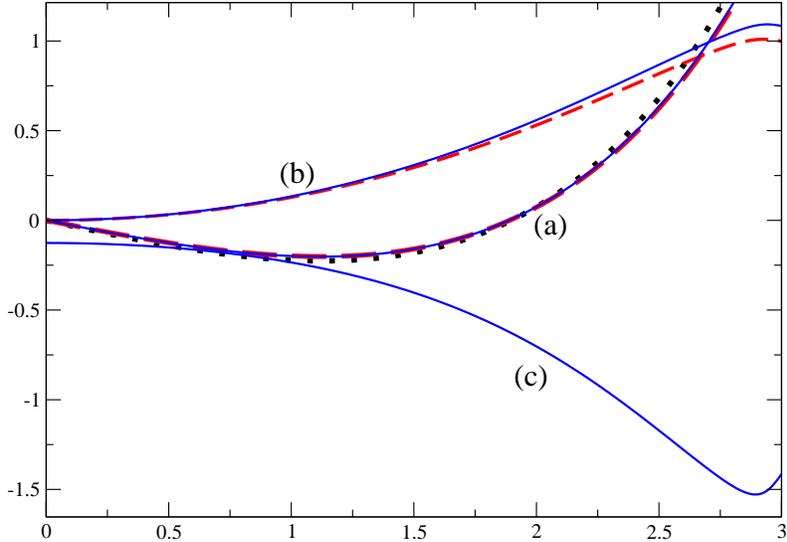}
\caption{\label{star}Wilson-Fisher fixed point in $D=3$ for (a) the
first derivative of the potential $v_*'(x)$, (b) the deviation of
the wave function renormalisation from its classical value $10\,
(z_*(x)-1)$, and (c) the four-derivative operator $10^3\, w_*(x)$.
Coding: local potential approximation (black dotted curve), 2nd
order derivative expansion (red dashed curves), 4th order derivative
expansion (blue continuous curves). }
\end{figure}

The results are displayed in Fig.~\ref{star}.  Including the wave
function renormalisation $z_*(x)$, the first derivative of the
potential $v'_*(x)$ changes only mildly from the local potential
approximation result.  The further inclusion of $w_* (x)$ leaves
$v_*'(x)$ practically unchanged, whereas the wave function
renormalisation $z_*(x)$ increases mildly, though only for larger
$x$. We note that $w_*(x)$ is very small and negative for small $x$,
enhancing its impact for smaller $x$. Some characteristic values of
the scaling solution are given in Tab.~\ref{results1}. Including
second (fourth) order operators, the vacuum expectation value changes
approximately by $1\%$ ($0.1\%)$, the curvature $v''$ by $10\%$
$(1.5\%)$, and the wave function renormalisation by $5\%$
$(0.5\%)$.\step

\begin{table}
\begin{center}
\begin{tabular}{ccccccc}
 \hline
  \hline
  approximation&
$\eta$ & $\nu$ & $\omega$ & $\omega_5$ & $\Delta$ &$\Delta_5$
\\ \hline
${}\quad$
LPA ${}\quad$ & 
$ 0$& $0.6260$ & $ 0.762$ & $2.163$ & $0.477$ & $1.354$\\
${}\quad$ ${\cal O}(\partial^2)$ ${}\quad$ &
$ 0.0330$  & $0.6244$ & $ 0.852$ & $2.459$ & $0.532$ & $1.535$\\
\ \ ${}\quad$ ${\cal O}(\partial^4)$
${}\quad$ \ \ 
& ${}\quad$ 0.0313 ${}\quad$ & ${}\quad$ 0.6247
${}\quad$ & ${}\quad$ 0.865 ${}\quad$ & ${}\quad$ 2.563 ${}\quad$&
${}\quad$ 0.540 ${}\quad$ & ${}\quad$ 1.601
${}\quad$\\
 \hline \hline
 \end{tabular}
\end{center}
${}$\vskip-.5cm \caption{\label{results2}Anomalous dimension, leading
  and sub-leading scaling exponents, and the Wegner corrections
  $\Delta=\nu\,\omega $ and $\Delta_5=\nu\,\omega_5$ for different
  orders in the derivative expansion (see text).}
\end{table}

\begin{figure}
${}$\vskip1cm \epsfig{file=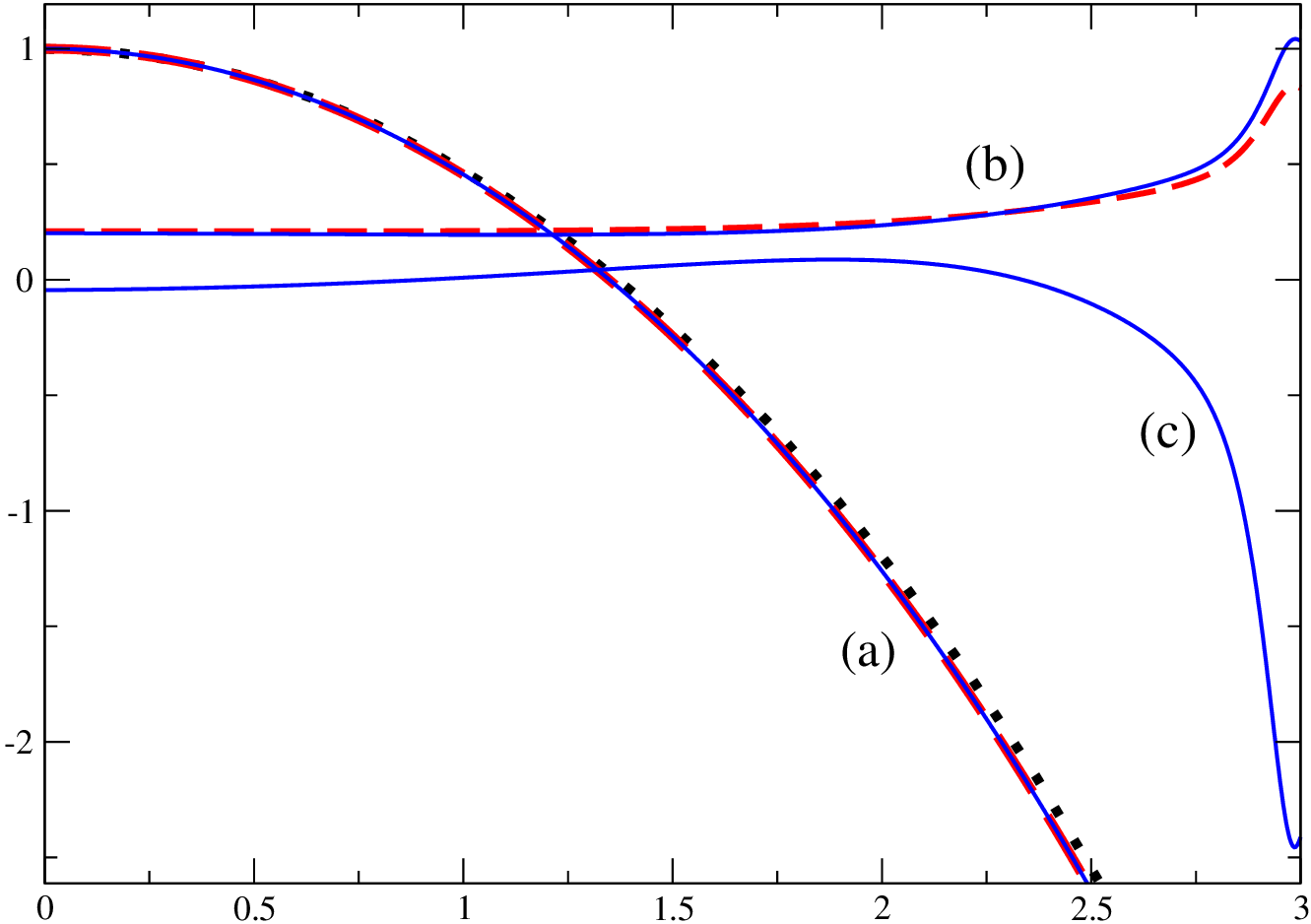,width=.65\hsize}
\caption{\label{linnu}Eigenperturbations $\Phi(x)$ with eigenvalue
$\nu$ at the Wilson-Fisher fixed point for (a) the potential
$v'_*(x)$, (b) $z_*(x)$ (rescaled by a factor $10 $), and for (c)
$w_*(x)$ (rescaled by a factor $10^3$). Coding as in
Fig.~\ref{star}.\\[2ex]}
\end{figure}
\begin{figure}
\epsfig{file=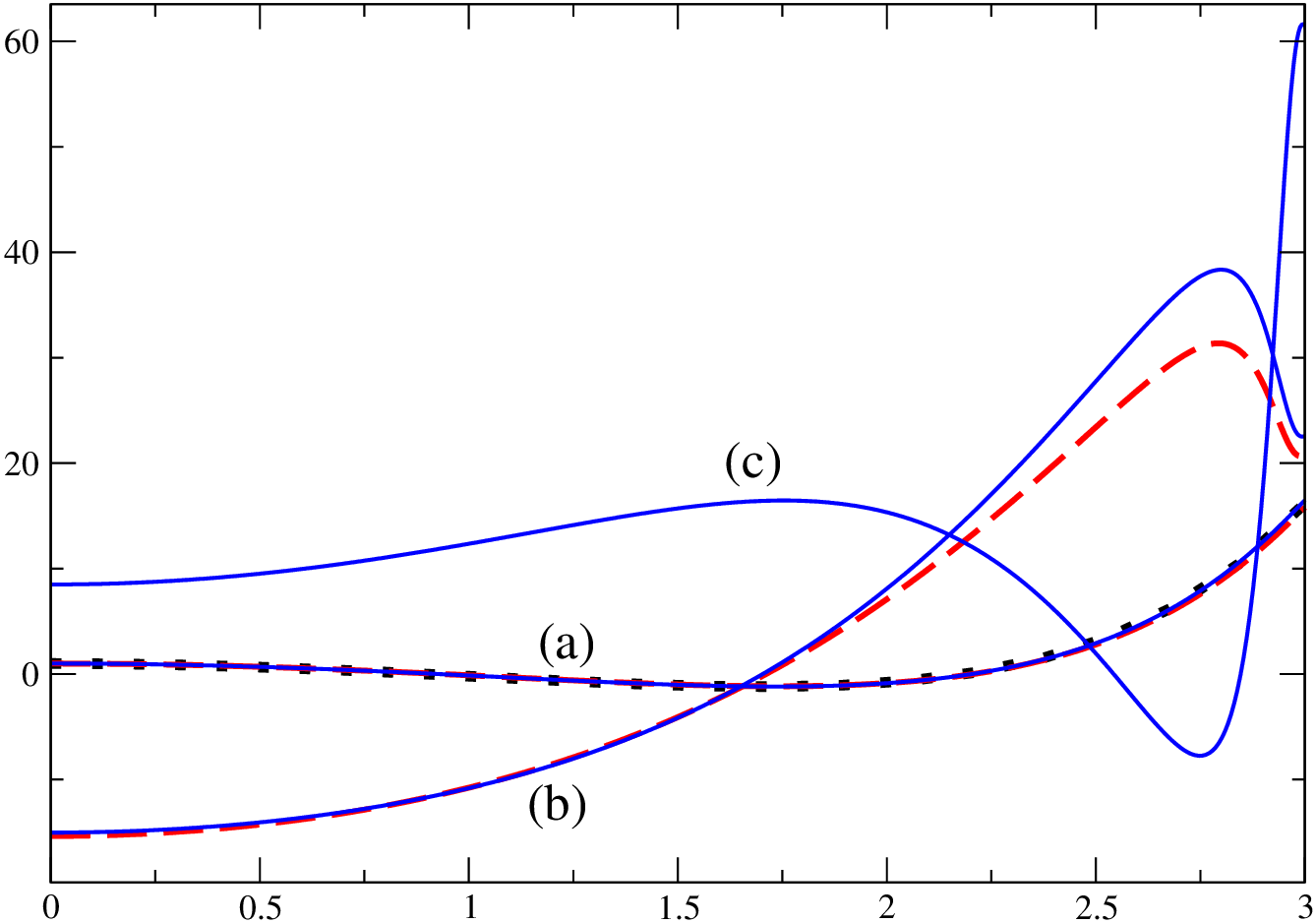,width=.65\hsize}
\caption{\label{linom}Eigenperturbations $\Phi(x)$with eigenvalue
$\omega$ at the Wilson-Fisher fixed point for (a) the potential
$v_*'(x)$, (b)  $z_*(x)$ (rescaled by a factor $10$), and for (c)
$w_*(x)$ (rescaled by a factor $10^3$). Coding as in
Fig.~\ref{star}.}
\end{figure}

We point out that the quantitative relevancy of operators in the
effective action correlates with their scaling dimension. This is
already visible from the results to lower orders in the derivative
expansion. We have $D_{u'}>0>D_{z}>D_{w}$, which materializes at the
fixed point \eq{FP} as variations in the scaling solutions $u_*'$,
$z_*$ and $w_*$ of order $1, 10^{-1}$, and $10^{-3}$, respectively
(see Fig.~\ref{star}).  Quantitatively, this can be understood as
follows. The exponential suppression of terms on the right-hand side
of \eq{zadim} for large field variable $x$ implies that the
large-field behaviour of operators is solely determined by their mass
dimension, see \eq{scaling}.  The transition from small-field
behaviour to large-field asymptotics is exponentially strong, thereby
restricting the impact of higher derivative operators to the small
field regime.  The same observation applies for the variation of the
fixed point solution and for the eigenperturbations at criticality
under successive extensions from $V(\phi)\to V(\phi), Z(\phi) \to
V(\phi),Z(\phi),W(\phi)$.\step

Next, we comment on the approximation \eq{eq:formgam}. A full ${\cal
  O}(\partial^4)$ order calculation in the derivative expansion of the
effective action requires the inclusion of the terms $H$ and $J$, see
\eq{eq:neglect}. Close to the free field theory limit, the three terms
in \eq{eq:neglect} scale identically, but this degeneracy is lifted at
a non-trivial fixed point solution. The mass dimensions of $W,H$ and
$J$ in \eq{eq:neglect} are different, and increasingly negative,
eg.~$D_w\equiv [W]=-(2+\eta)$, $D_{h}\equiv [H]=-\s012(5+3\eta)$ and
$D_{j}\equiv[J]=-(3+2\eta)$ in $D=3$ dimensions.  Hence
$0>D_w>D_{h}>D_{j}$ and, consequently, the scaling solutions $h_*(x)$
and $j_*(x)$ will be suppressed compared to $w_*(x)$, see
\eq{scaling}. Therefore we expect that the impact of $H$ and $J$ on
scaling exponents is subleading, analogous to the pattern observed to
lower orders in the derivative expansion. We also note that the
suppression, in general, will depend quantitatively on the
regularisation. The suppression is exponential for the background
field flow used here, and hence stronger than the power-law
suppression observed for standard flows in a derivative expansion, see
\cite{Ballhausen:2003bu,Canet:2002gs,Canet:2003qd}. \step

Small deviations from the fixed point
$\Phi_{Y,\lambda}(x)=Y(x)-Y_*(x)$ are classified according to their
universal scaling exponents $\lambda$. In the vicinity of the fixed
point the eigenperturbations obey the eigenvalue equation
\be\label{EW}
\partial_t \Phi_{Y,\lambda} = \lambda\,\Phi_{Y,\lambda}\,.
\ee 
We solve \eq{EW} using \eq{zadim} and the fixed point solution \eq{FP}
to find the leading and subleading eigenvalues as well as the
eigensolutions. The leading eigenperturbations $\Phi(x)$ are symmetric
under $x\leftrightarrow -x$. The eigenvalues obey
$\lambda_0<0<\lambda_1<\lambda_2<\cdots$ with
$\-\lambda_0\equiv-1/\nu$ and $\lambda_1\equiv \omega$ in the
statistical physics literature. The eigenperturbations $\Phi(x)$ which
are antisymmetric under under $x\leftrightarrow -x$ have eigenvalues
$0<\bar\lambda_1<\bar\lambda_2<\cdots$, and the smallest eigenvalue is
denoted as $\bar\lambda_1\equiv \omega_5$ in the literature; see
\cite{Litim:2003kf} for a determination of $\omega_5$ in the local
potential approximation.  Our results for the eigenperturbations are
given in Figs.~\ref{linnu} and~\ref{linom}. The inclusion of $z_*(x)$
and $w_*(x)$ changes the eigenperturbations with eigenvalue $\nu$
($\omega$) only mildly from the local potential approximation.  For
the scaling exponents and the anomalous dimension, we find
\begin{equation}\label{4th}
\nu=0.6247\,,\quad
\eta=0.0313\,,\quad
\omega=0.865\,,\quad
\omega_5= 2.563\,.
\end{equation}
The numerical precision for $\omega$ $(\omega_5)$ is of the order
$0.1\%$ $(1\%)$.  Comparing with lower orders in the derivative
expansion, Tab.~\ref{results2}, we conclude that the derivative
expansion of the background field flow displays a very good numerical
convergence. \\

\begin{table}
\begin{center}
\begin{tabular}{llllllll}
 \hline \hline
${}_{}\quad\quad\quad$info &
${}\quad$ $\gamma$ &
${}\quad$  $\nu$ &
${}\quad$  $\eta$ &
${}\quad$ $\alpha$& 
${}\quad$ $\beta$ & 
${}\quad$  $\delta$     & 
${}\quad$  $\omega$ \\
a) world average\ \ &
1.2372(5) \ \ &0.6301(4) \ \ & 0.0364(5) \ \ 
& 0.110(1) \ \ & 0.3265(3) \ \ & 4.789(2) \ \ & 0.84(4)  \\
b) this work & 
1.2298& 0.6247 & 0.0313 & 0.1259& 0.3221& 4.818& 0.865
\\
c) 
impl.~opt.&
 1.233 & 0.627 & 0.034 & 0.119& 0.324& 4.803& 0.839
\\ \hline \hline
 \end{tabular}
\end{center}
\caption{\label{results3}Comparison of a) the world average of
  theoretical predictions \cite{Pelissetto:2000ek}, with our results b)  \eq{4th} 
  and c) \eq{opt4}, also using scaling and hyper-scaling relations.}
\end{table}

In Tab.~\ref{results3}a) and b), we compare the `world average' of
scaling exponents in three dimensions as compiled in
\cite{Pelissetto:2000ek} with our findings \eq{4th}.  Most exponents
agree on the percent level and below, in particular the indices $\nu$,
$\gamma$, $\delta$ and $\beta$ which are predominantly sensitive to
the field-dependence of vertices at vanishing momentum.  The anomalous
dimension $\eta$, and the exponents $\alpha$ and $\omega$ are
subleading and more sensitive to the momentum structure of propagators
and vertices. Consequently, their precision is lower.  Interestingly,
the exponent $\omega$ already agrees with the world average within
$3\%$. The exponent $\alpha$ and the anomalous dimension $\eta$ only
agree within $15\%$ with the best values quoted in the literature.
The same pattern persists the comparison with recent high accuracy
results from Monte Carlo simulations $\nu_{\rm MC}=0.63002(10)$,
$\omega_{\rm MC}=0.832(6)$, $\eta_{\rm MC}=0.3627(10)$
\cite{Hasenbusch:2010}.  Here, the indices
$\nu,\Delta=\nu\,\omega,\omega$ and $\eta$ agree to within $0.8\%,
3\%, 4\%$ and $13\%$, respectively.\\

\noindent{\bf 5. Optimisation}\step

In this section, we discuss the optimisation of results based on a
background field flow. It is well-known that physical observables
within an approximation of the functional flow can depend on the shape
of the momentum cutoff function $R$ and its parameters,
eg.~$m$. The reason for this is that the cutoff function, a non-trivial
 function of momenta, couples to all operators in the effective action. 
 Neglecting some operators means that some cutoff-dependent 
 back-coupling in \eq{flow} is missing.
Formally, within some systematic expansion of the flow
equation, one obtains the exponents as a series
\begin{equation}\label{expansion}
\nu_{\rm phys}=\nu_{(0)}(R)+\nu_{(1)}(R)+\nu_{(2)}(R)+\cdots\,,
\end{equation}
where the contribution from every single order $\nu_{(n)}(R)$ may
depend on the cutoff function $R$, and only the full physical result
will be independent thereof.  Optimisation is based on the observation
that the convergence of \eq{expansion}, and similarly for other
observables, is improved through optimised choices for $R$, $ie.$~the
parameter $m$. \step

To evaluate the $R$-dependence of our results -- and also as a
consistency check for our numerical codes -- we have re-calculated the
exponents to leading and second order in the derivative expansion
using \eq{PT} for other values of the cutoff parameter $m$
\cite{Bonanno:2000yp,Mazza:2001bp,Litim:2001hk}. The output is
displayed in Tab.~\ref{resultsms} and, as expected, the results fully
agree with earlier findings.  We add the following observations. To
leading order in the derivative expansion, $\nu(m)$ is monotonous,
covering the physical value $\nu_{\rm av}\approx0.63$. This is
different from the pattern observed using the standard flow in the
same approximation \cite{Litim:2002cf}.  The index $\omega(m)$ is
equally monotonous, approaching the world average value $\om_{\rm
  av}\approx 0.84$ from below, $\om(m)<\om_{\rm av}$. Together with
the improved convergence of the amplitude expansion, this justifies
the use of $1/m=0$, also favoured by a minimum sensitivity
condition. We note that the physical value for $\nu$ is matched at
$1/m\approx 0.11$.  \step

To second order in the derivative expansion, all three observables
$\nu$, $\omega$ and $\eta$ are monotonous functions of $m$. The ranges
covered include the physical value in all three cases. A principle of
minimum sensitivity is not applicable.  For $\nu$ $(\omega)$, the
relative change from leading to second order is minimal at $1/m=0$
$(1/m\approx0.4)$.  We can use our results to second order in the
derivative expansion to identify the value of the cutoff parameter $m=
m_{\rm av}$ for which $\eta$ -- the least well-determined index --
matches best with the prediction from the `world average' or Monte
Carlo studies. We find $1/m_{\rm av}\approx 0.08$, which is very 
close to the best match $1/m\approx 0.11$ found to leading order, and
\begin{equation}\label{eta2nd}
\nu=0.626\,,\quad
\eta=0.036\,,\quad
\omega=0.823\,. 
\end{equation}
The difference between \eq{eta2nd} and Monte-Carlo results for $\nu$,
$\omega$ and $\Delta$ reduces to $0.7\%$, $1.6\%$ and $2.3\%$,
respectively. The improved agreement shows that the scaling exponents
display the correct cross-dependences, also supporting small values
for the inverse cutoff parameter $1/m$.  The fit $ {1}/{m_{\rm
    av}}\approx 0.08$ comes out slightly larger than ${1}/{m}=0$, a
consequence of the anomalous dimension being underestimated in the
latter case.  The cross-correlation amongst $\nu$, $\omega$ and $\eta$
as functions of the cutoff parameter $m$ is similar to the strong
cross-correlation observed earlier in the local potential
approximation \cite{Litim:2007jb}. \step
\begin{table}
\begin{center}
\begin{tabular}{lccccc}
\hline \hline ${}$\quad$m$& ${}\quad 2\quad$ & $\quad 2.5\quad$ 
& $\quad 3\quad$ & $\quad 4 \quad$& $\quad \infty \quad$
\\ \hline
${}\quad \eta|_{{\cal O}(\partial^2)}$ ${}\quad$ & $0.065$ & $0.056$
& $0.051$ & $0.045$& $0.033$
\\ \hline
${}\quad \nu|_{\rm LPA}$ ${}\quad$ & $0.660$ & $0.650$ & $0.644$
 &
$0.638$ &$0.626$ 
\\
${}\quad \nu|_{{\cal O}(\partial^2)}$ ${}\quad$ & $0.632$ &
$0.631$ & $0.630$ & $0.629$& $0.624$
\\ \hline
${}\quad \omega|_{\rm LPA}$ ${}\quad$ & $0.628$ & $0.656$ &
$0.674$
 &
$0.698$&$0.762 $
\\
${}\quad \omega|_{{\cal O}(\partial^2)}$ ${}\quad$ & $0.677$ &
$0.702$ & $0.725$ & $0.756$ & $0.852$ 
\\  \hline \hline
 \end{tabular}
\end{center}
${}$\vskip-.5cm \caption{\label{resultsms} Variation of scaling
  exponents with cutoff shape parameter $m$; background field flow
  \eq{PT} to leading and second order in the derivative expansion.}
\end{table}

Using the above, we obtain an improved estimate to fourth order in the
derivative expansion by combining \eq{4th} with the $m$-dependence of
indices for small $1/m$ from Tab.~\ref{resultsms} at $1/m\approx
0.08$.  Note that at large $m$, the difference $m\frac{{\rm d}X}{{\rm
    d}m}|_{O(\partial^2)}-m\frac{{\rm d}X}{{\rm d}m}|_{O(\partial^4)}$
in the $m$-dependence of exponents with $X=\nu,\omega$ or $\eta$
between the 2nd order and 4th order results are small. With this approximation, 
we arrive at
\begin{equation}\label{opt4}
\nu=0.627\,,\quad
\eta=0.034\,,\quad
\omega=0.839\,.
\end{equation}
The difference between \eq{4th} and \eq{opt4} serves as a measure for
the variation of indices within the stable domain of RG flows to this
order in the approximation. In Tab.~\ref{results3}c), we compile our
findings \eq{opt4} and compare with the world average.  The agreement
with Tab.~\ref{results3}a) becomes significantly enhanced. For the
indices $\gamma$, $\nu$, $\eta$, $\alpha$, $\beta$, $\delta$,
$\Delta=\omega\,\nu$ and $\omega$ we find an accuracy of $3\%,0.5\%,
7\%, 8\%,0.3\%, 0.2\%$ and $0.6\%$, respectively.  The same quality
persists the comparison with recent high accuracy results from Monte
Carlo simulations \cite{Hasenbusch:2010}.  Here, the indices
$\nu,\Delta=\nu\,\omega,\omega$ and $\eta$ agree to within $0.5\%,
0.4\%, 0.9\%$ and $6\%$, respectively, which is a clear improvement
over the results at $1/m=0$ (see Sec.~4).\step

In summary, optimisation of background field flows consistently
favours small values for the parameter $1/m$. The excellent agreement
of indices around $1/m\approx 0.08$ indicates that the operators
retained in our approximation display the physically expected
cross-correlations.  This non-trivial result lends additional support
for the present set-up and the internal consistency of the underlying
approximations.\\

\noindent{\bf 6. Variation with dimensionality}\step

In this section, we consider the Wilson-Fisher fixed point away from
three dimensions.  It is a useful consistency check to understand the
global $D$-dependence of our findings, and their interpolation between
the known results in two and four dimensions. Furthermore, probing the
local $D$-dependence by perturbing the system with $\frac{\rm d}{{\rm
    d}D}|_{D=3}$ provides insights into the structural stability of
our set-up. We note that this information is also of interest for
systems of finite size or finite volume, where variations of the
(spatial) dimensions are sensitive to the variations with linear
system size.\step

With increasing dimensionality $D>3$, the scaling exponents approach
mean field values at $D=4$ with $\eta_{4d}=0$, $\nu_{4d}=\frac{1}{2}$
and $\omega_{4d}=0$.  For decreasing $D<3$, further higher-order
critical points become accessible whenever $n\approx D/(D-2)$ becomes
an integer, with $n=3$ corresponding to the Wilson-Fisher fixed point.
In $D=2$, scaling exponents and the anomalous dimension take the known
values $\nu_{2d}=1$, $\omega_{2d}=1$ and $\eta_{2d}=\s014$. \step

We have computed $\nu$, $\omega$ and $\eta$ in the vicinity of three
dimensions, see Tab.~\ref{results4}. Below $D\klgl 2.7$, the
identification of the scaling solution becomes numerically more
demanding. This should be related to the appearance of a competing
scaling solution, which becomes available when $D\approx 8/3$.  From
the data, we find
\begin{equation}\label{Dexp}
\frac{{\rm d}\nu}{{\rm d}D} =-0.18\,,\quad \frac{{\rm d}\omega}{{\rm d}D}
=-0.65\,,\quad \frac{{\rm d}\eta}{{\rm d}D} =-0.08\,.\quad
\end{equation} 
for the first derivatives at $D=3$. Note that the derivatives would
read $-0.25,-0.5$ and $-0.125$ for a simple linear interpolation
between the analytically known results at $D=2$ and 4.  Fitting the
data points for $\nu$ and $\eta$ ($\omega$) with a cubic (quadratic)
polynomial in $D$, and extrapolating we find
$\nu|_{D=4}\approx0.49$, $\omega|_{D=4}\approx 0.02$ and
$\eta|_{D=4}\approx-0.02$. In the opposite limit, extrapolation leads
to $\nu|_{D=2}\approx 0.92$, $\omega|_{D=2}\approx 1.3$, and
$\eta|_{D=2}\approx 0.2$. These estimates are fully consistent with
the expected behaviour, and the slight deviations at the endpoints
serve as (rough) indicator for the underlying error. The extrapolated
result for $\eta$ is smaller than the exact one $\eta=\frac{1}{4}$,
and the fourth order result is slightly smaller than the second order
result. This suggests that our result slightly under-estimates the
value for $\eta$ at $D=3$, though a definite conclusion would require
more data points for $\eta$ in $2<D<3$. For a study of $\eta$ and 
$\nu$ in $1<D<4$ dimensions using an optimised standard
flow to second order in the derivative expansion, we refer to
\cite{Ballhausen:2003gx}.\step

Next, we estimate the relative variation of scaling exponents. Suppose
we are interested in a physical observable $X$. The relative variation
of $X$ with $D$ around the dimensionality of interest serves as an
indicator for the stability in the observable $X$. A low stability
indicates that an observable will depend more strongly on the
approximation, and vice versa. With this in mind, we write ${\rm d}
X/{\rm d}D=-A_X\, X$ for $X=\nu,\omega,\eta$. Using \eq{Dexp}, we find
\begin{equation}\label{A}
A_\nu= 0.28(1)\,,\quad A_\omega=0.85(1)\,,\quad A_\eta=2.5(1)
\end{equation}
in three dimensions. The smallness of $A_\nu$ explains the high
accuracy achieved for $\nu$ already to low orders in a derivative
expansion. In turn, the large value of $A_\eta$ explains the stronger
sensitivity of $\eta$ on the approximation. Furthermore, the pattern
$A_\nu<A_\omega<A_\eta$ suggests that the expected accuracy in
$\omega$ should be better than the one in $\eta$, and worse than the
one in $\nu$. This is in accord with the pattern observed in our
results, see Sect.~4, and with the earlier functional RG results
discussed below (see Sect.~8 and 9).\\

\begin{table}
\begin{center}
\begin{tabular}{cllcccc}
\hline\hline
\quad $D\quad$
&  ${}\quad$$\nu|_{{\cal O}(\partial^0)}$$\quad$
&  ${}\quad$$\omega|_{{\cal O}(\partial^0)}$$\quad$
&  ${}\quad$$\eta|_{{\cal O}(\partial^2)}$$\quad$
 & ${}\quad$$\eta|_{{\cal O}(\partial^4)}$${}\quad$
\\
\hline
3.3&0.577915&0.559475&&\\
3.2&0.592702&0.628553&&\\
3.1&0.608674 &0.696103&&\\
3.0&0.625979&0.762204 &0.0330   & 0.0313 \\
2.9    &0.644808&0.82685   &0.0418   & 0.0400 \\
2.8    &0.665407& 0.8899    &0.0519   & 0.0502 \\
2.7    &0.688&0.949    &0.0637   & 0.0621 \\
\hline\hline
\end{tabular}
\end{center}
\vskip-.5cm \caption{\label{results4}Variation of $\nu$, $\omega$ and
  $\eta$ with dimensionality $D$ to leading and second order in the
  derivative expansion (see text). }
\end{table}

\noindent{\bf 7. Cross-correlations}\step

Cross-correlations amongst scaling exponents provide insights into the
finer structure of the theory, and into the inner working of the
approximation in place.  Within the local potential approximation,
cross-correlations are strong \cite{Litim:2007jb}, and only weakly
dependent on the cutoff $R_k$, in particular for optimised flows
\cite{Litim:2000ci}. A similar cross-correlation has been observed
based on hierarchical RG transformations, thereby providing a link
between the cutoff $(R_k)$ dependence of the continuum RG and finite
step size effects in discrete versions thereof \cite{Litim:2007jb}.
\step

Here, we are interested in the correlations to higher order in the
derivative expansion.  To set the stage, we perform a linear
interpolation for the derivatives based on the known results at $D=2$
and $D=4$. We find
\begin{equation}\label{cc-lin}
\frac{{\rm d}\omega}{{\rm d}\nu}
={2}\,,\quad
\frac{{\rm d}\eta}{{\rm d}\nu}
= 0.5\,,\quad 
\frac{{\rm d}\eta}{{\rm d}\omega}
=0.25\,.
\end{equation}
Within our functional RG set-up, we access the cross-correlation of
exponents by keeping the regulator fixed, and by exploiting that
\eq{Dexp} represent full variations with $D$.  Since $\eta(D)$ is
monotonous in $D$, at least in the region of interest (see
Tab.~\ref{results4}), we invert $\eta(D)$ into $D(\eta)$ to obtain the
functions $\nu(\eta)\equiv \nu(D(\eta))$ and $\omega(\eta)\equiv
\omega(D(\eta))$ which encode the cross-correlation of scaling
exponents. In three dimensions, their first derivatives read
\begin{equation}\label{cc}
\frac{{\rm d}\omega}{{\rm d}\nu}
=3.63\,,\quad
\frac{{\rm d}\eta}{{\rm d}\nu}
=0.45\,,\quad 
\frac{{\rm d}\eta}{{\rm d}\omega}
=0.124\,.
\end{equation}
Note that $\frac{{\rm d}\omega}{{\rm d}\nu}\frac{{\rm d}\nu}{{\rm
    d}\eta}\frac{{\rm d}\eta}{{\rm d}\omega}=1$ to within $0.03\%\,$,
which is smaller than the error in \eq{cc}. Comparing \eq{cc} with the
linear approximation \eq{cc-lin}, we find that $ \frac{{\rm
    d}\eta}{{\rm d}\nu}$ is roughly of the same size, while
$\frac{{\rm d}\omega}{{\rm d}\nu}$ ($\frac{{\rm d}\eta}{{\rm
    d}\omega}$) is roughly twice (half) as big as the linear
approximation. Our result \eq{cc} compares well with the estimate
$\frac{{\rm d}\eta}{{\rm d}\nu}|_{\eps-{\rm exp.}}=0.59$ obtained from
a modified epsilon expansion \cite{Guida:1998bx}. We note that
$\frac{{\rm d}\eta}{{\rm d}\nu}|_{\rm fRG}<\frac{{\rm d}\eta}{{\rm
    d}\nu}|_{\rm lin.}<\frac{{\rm d}\eta}{{\rm d}\nu}|_{\eps-{\rm
    exp.}}$. The double-logarithmic derivatives follow from \eq{A} in
an obvious manner, eg.~$\frac{{\rm d}\ln \omega}{{\rm d}\ln \nu}\equiv
A_\omega/A_\nu$, leading to the estimates $\frac{{\rm d}\ln
  \omega}{{\rm d}\ln \nu} \approx 3\,, \frac{{\rm d}\ln \eta}{{\rm
    d}\ln \nu} \approx 9$ and $\frac{{\rm d}\ln \eta}{{\rm d}\ln
  \omega} \approx 3$, consistent with the sensitivity observed in our
results.\\

\noindent{\bf 8. Convergence}\step

Despite the small anomalous dimension, the Wilson-Fisher fixed point
corresponds to a non-trivially interacting theory and is therefore
intrinsically non-perturbative. While little is known about the
absolute convergence of systematic approximations to \eq{flow} in the
non-perturbative regime, the numerical convergence of expansions can
be accessed order by order \cite{Litim:2001dt}.  In this section, we
discuss the convergence of the derivative expansion (see
Tab.~\ref{DE}) by comparing results for $\nu$, $\eta$ and $\omega$
from different realizations of the functional renormalisation group
including the standard flow \eq{flow}, background field flows \eq{PT},
and the Wilson-Polchinski flow (see \cite{Litim:2001dt} for an earlier
overview). We have omitted data points which are not based on an (at
least partly) optimised choice for the momentum cutoff, eg.~sharp
cut-off results \cite{Hasenfratz:1985dm}.  \step

To leading order in the derivative expansion, the full cutoff
dependence of $\nu(R)$ is known within the standard flow
\cite{Litim:2002cf,Litim:2007jb}, within the Wilson-Polchinski flow
\cite{Litim:2005us} where the result is unique, and, partly, within
background field flows
\cite{Bohr:2000gp,Bonanno:2000yp,Mazza:2001bp,Litim:2001hk}. For the
standard flow, the best result is given in b) \cite{Litim:2002cf},
achieved for suitably optimised regulators.  High-accuracy expressions
for the exponents are given in \cite{Bervillier:2007rc} and are in
full agreement with findings from the Wilson-Polchinski flow
\cite{Litim:2005us} in c).  The background field flow covers a larger
range of values for $\nu(m)$, the smallest one given in a). Comparing
a) with b), we note that the leading index $\nu$ (subleading index
$\omega$) is slightly (significantly) closer to the physical result in
the setup a).\step

For the ${\cal O}(\partial^2)$ approximation, we report the exponents
from the standard flow based on an optimised algebraic (power-law)
cutoffs \cite{Morris:1994ie} in d), 
a standard exponential cutoff \cite{Seide:1998ir,VonGersdorff:2000kp}
in e), an optimised exponential cutoff \cite{Canet:2002gs} in f), and a
flat optimised cutoff \cite{Canet:2002gs} in g). Note that algebraic
(power-law) cutoffs of \cite{Morris:1994ie} leads to slowly converging
flows within the derivative expansion \cite{Litim:2001dt}, which is
already visible within the local potential approximation
\cite{Litim:2002cf}. The comparatively large estimate for $\eta$ in e)
is a consequence thereof. Below, we will retain e) for a conservative
error estimate.  Comparing d)-g) with h), we note that the indices
$\nu$ and $\omega$ differ only slightly amongst the different
implementations. In contrast, the anomalous dimension $\eta$ varies
more strongly, about $\pm25\%$.  In the standard flow, the anomalous
dimension stays above $4\%$, whereas the background field flow leads
to a result below $4\%$, closer to the physical value.\step

Results to second order in the derivative expansion are also available
within Polchinski's formulation of the renormalisation group
\cite{Polchinski:1983gv,Bervillier:2005za}.  The Wilson-Polchinski
flow is linked to \eq{flow} by a Legendre transform, implying that
derivative expansions are inequivalent beyond the trivial order.  A
significant cutoff dependence, in particular for $\eta$, is observed
\cite{Ball:1994ji,Comellas:1997ep,Bervillier:2005za}, which calls for
a stability-based optimisation of the cutoff
\cite{Litim:2000ci,Litim:2005us,Pawlowski:2005xe}.  A prediction for
Ising exponents is achieved at $O(\partial^2)$ by tuning the cutoff to
the desired value for $\eta$, say $\eta\approx 0.038$
\cite{Ball:1994ji}, and using a minimum sensitivity condition to
identify the remaining exponents (see also
\cite{Comellas:1997ep,Bervillier:2005za}). This leads to $\nu\approx
0.625$ and $\omega\approx 0.77$ \cite{Ball:1994ji}, summarized in
Tab.~\ref{DE}i). The predictions for $\nu$ and $\omega$ are in the
expected range of values, showing that the Wilson-Polchinski flow
displays the correct cross-correlation of scaling exponents. It will
be interesting to see whether a fourth-order computation stabilises
the result.  For comparison, we have added in Tab.~\ref{DE}j) our
result \eq{eta2nd} from the background field flow to second order in
the derivative expansion, where $\eta$ has been matched to the world
average and Monte-Carlo result. The Wilson-Polchinski and background
field estimates agree very well for $\nu$, and differ by less than
$8\%$ for the exponent $\omega$. The background field value is much
closer to the expected value.  Note that this procedure is not
applicable for the standard flow to second order, because the
anomalous dimension stays above $4\%$ for all cutoffs and cannot be
matched to the physical value.\step

\begin{table}[t]
\begin{center}
\begin{tabular}{clllll}
\hline \hline & ${}\quad$ info & $\quad \quad \eta$ & $\quad \quad
\nu$ & $\quad\quad \omega$ & \quad\quad refs.
\\
\hline 
a)&\quad $\partial^0$, bf 
& \quad $0$ 
&\quad $0.6260\ \  \,\,$ 
&\quad $0.7622\ \ \ \,$   
& \quad  \cite{Mazza:2001bp,Litim:2001hk}, this work
\\
b)&\quad $\partial^0$, st 
&\quad $0$ 
&\quad $0.649\,561\cdots$ 
&\quad $0.655\,746\cdots\ \ $  
& \quad \cite{Litim:2001dt,Litim:2002cf,Bervillier:2007rc}
\\
c)&\quad $\partial^0$, WP
&\quad $0$ 
&\quad $0.649\,561\cdots$ 
&\quad $0.655\,746\cdots\ \ $  
& \quad \cite{Litim:2005us,Bervillier:2007rc}
\\
d)&\quad $\partial^2$, st, alg  
&\quad $0.05393$
&\quad$0.6181\ \,$ 
&\quad $0.8975$ 
& \quad \cite{Morris:1994ie}
\\
e)&\quad $\partial^2$, st, exp 
&\quad $0.0467\ $ 
&\quad $0.6307\ $
&\quad $-$ 
& \quad\cite{Seide:1998ir,VonGersdorff:2000kp}
\\
f)&\quad $\partial^2$, st, exp, opt 
&\quad $0.0443\ $ 
&\quad $0.6281\ $
&\quad $-$ 
& \quad\cite{Canet:2002gs}
\\
g)&\quad $\partial^2$, st, opt 
&\quad $0.0470\ $ 
&\quad $0.6260\ $
&\quad $-$ 
& \quad\cite{Canet:2002gs}
\\
h)&\quad $\partial^2$, bf 
&\quad $0.0330\ $ 
&\quad $0.624\ \ \,$
&\quad $0.852\ \ \ \,$ 
& \quad\cite{Mazza:2001bp}, this work
\\
i)&\quad $\partial^2$, WP, $\eta$-matching
&\quad $0.038$
&\quad $0.625$
&\quad $0.77$
&\quad\cite{Ball:1994ji}
\\
j)&\quad $\partial^2$, bf, \ \ $\eta$-matching
&\quad $0.036$
&\quad $0.626$
&\quad $0.823$
&\quad this work\\
k) &\quad $\partial^4$, st, exp, opt
&\quad 0.033\ \ \, 
&\quad 0.632\ \ \  \,\,
&\quad $-$ 
& \quad\cite{Canet:2003qd}
\\
l)&\quad $\partial^4$, bf 
 &\quad $0.0313\ $ 
 &\quad $0.6247\ \ \,$ 
 &\quad  $0.865\ \ \ \,$ 
 & \quad this work
\\
m)&\quad $\partial^4$, bf, implicit
&\quad $0.034\ \ \,$ 
&\quad $0.627\ \ \ \,\,$ 
&\quad $0.839\ \ \ \, \ \,$ 
& \quad this work
\\
n)&\quad mixed, st, exp, opt
&\quad $0.039\ \ \,$ 
&\quad $0.632\ \ \ \,\,$ 
&\quad $0.78\ \ \ \, \ \,$ 
& \quad \cite{Benitez:2009xg}
\\
\hline \hline
\end{tabular}
\end{center}
\caption{\label{DE} Comparison of results from 
the functional renormalisation group within various approximations 
(see text). 
Local potential approximation ($\partial^0$):
a) background field flow (bf); 
b) standard flow (st);
c) Wilson-Polchinski flow (WP).
Derivative expansion to second order ($\partial^2$): 
d) - g) standard flow (various cutoffs);
h) background field flow.  
Derivative expansion to second order with matching of the 
anomalous dimension:
i) Wilson-Polchinski flow;
j) background field flow. 
Derivative expansion to fourth order ($\partial^4$): 
k) standard flow; l) background field flow; 
m) background field flow with implicit optimisation. 
Mixed approximation retaining momentum- and 
field dependences (mixed): n) standard flow.}
\end{table}

Beyond ${\cal O}(\partial^2)$ in the derivative expansion, we cite the
fourth order computation by \cite{Canet:2003qd} in k), which is
compared with our result \eq{4th} in l), the optimised background
field result \eq{opt4} in m) and the `mixed' analysis of
\cite{Benitez:2009xg} in n), which is optimised using the principle of
minimum sensitivity
\cite{Stevenson:1981vj,Litim:2001fd,Liao:1999sh}. The approach n)
retains momentum- and field-dependences in the ansatz for the
effective action, amended by approximations on the level of the flow;
see \cite{Blaizot:2005xy} for technical details. The results for $\nu$
in all approaches are very close to the world average $\nu_{\rm
  av}=0.6301(4)$. The value for $\omega$ from background field flows
l) and m) are closest to the world average $\omega_{\rm
  av}=0.84(4)$. All values for $\eta$ are now below $4\%$. Still, a
slight variation of $\eta$ remains visible which makes the anomalous
dimension the least well-determined obervable in Tab.~\ref{DE}.  We
note that the prediction for $\eta$ based on k) and n) are equally
close to the world average $\eta_{\rm av}=0.0362(4)$, approaching it
from opposite sides. This is interesting because n) should have a
better access to the momentum dependence of propagators.  We suspect
that the approximations on the level of the flow exercised in
\cite{Blaizot:2005xy} are responsible for this pattern. The
$\eta$-values from background field flows approach the physical value
from below, with m) being closest to the expected value.\step

The mean values based on all data points in Tab.~\ref{DE} are $\langle
\nu\rangle_{\rm fRG}=0.630$ and $\langle\omega\rangle_{\rm
  fRG}=0.790$. For the anomalous dimension, we find $\langle
\eta\rangle_{\rm fRG}=0.0312\, (0.0397)$, depending on whether we
retain (suppress) the LPA data points $\eta_{\rm LPA}=0$.  (We come
back to a detailed discussion of mean values and systematic errors in
Sect.~9.)  \step

We use the numerical convergence of the derivative expansion for a
crude error estimate. For the standard flow with order-by-order
optimised exponential cutoff function $R_k(q^2)\propto\alpha\,
q^2/(\exp q^2/k^2-1)$ we compare the LPA result $\eta=0$ and
$\nu=0.6506$ \cite{Litim:2000ci} with higher orders in the derivative
expansion Tab.~\ref{DE}f) and k). This leads to $\nu\approx 0.637$
$\pm 2\%$ and $\eta\approx 0.0387$ $\pm 15\%$. Retaining only the two
best values for $\nu$ improves the error estimate, $\nu\approx 0.630$
$\pm 0.3\%$. The relative change $\Delta\nu/\nu$ reads $3.5\times
10^{-2}$ ($6.3\times 10^{-3}$) at second (fourth) order in the
derivative expansion. For the background field flow with cutoff
$m\to\infty$, we compare Tab.~\ref{DE}a), h) and l), leading to
$\nu\approx 0.625$ $\pm 0.4\%$ and $\eta=0.0322\pm2\%$. Hence, in the
approximation \eq{PT}, the numerical convergence of background field
flows is slightly faster.\step

We conclude that the derivative expansion of the functional
renormalisation group, together with suitably optimised regulators,
shows a very good numerical convergence up to high order for both
standard and background field flows. \\

\noindent{\bf 9. Systematic errors}\step

Estimating systematic errors is common practice in eg.~lattice
approaches and resummations of perturbation theory. Here we discuss
how analogous estimates can be achieved for the functional
renormalisation group, where physical observables are obtained by
projecting the full flow in `theory space' -- the infinite dimensional
space of operators parametrizing the effective action $\Gamma_k$ --
onto a sub-set thereof. This step implies an approximation and is a
potential source for systematic errors. The flexibility of the
formalism, however, allows for many different projections. Then the
quantitative comparison of different projections gives access to the
systematic uncertainty.\step

We recall that, in general, approximations to the flow equation
\eq{flow} enter via operators neglected in the effective action
$\Gamma_k$, approximations on the level of the flow $\partial_t
\Gamma_k$, and the choice for the momentum cutoff $R_k$.  These
aspects are partly intertwined, to the least because a momentum cutoff
introduces a non-trivial momentum structure into the flow. In general,
the operator content is central. A similar importance should be given
to approximations on the level of the flow $\partial_t \Gamma_k$,
which feed back into the determination of scaling exponents.  The
regulator is crucial for the stability and convergence of the RG flow
\cite{Litim:2000ci}. Within given approximations for $\Gamma_k$ and
$\partial_t\Gamma_k$, the regulator can be optimised to maximise the
physics content in the flow, and to minimize cutoff artefacts.
Uncertainties due to the boundary condition for the effective action
are irrelevant for fixed point solutions. We conclude that systematic
errors should only be derived from `cutoff-optimised' results to
eliminate cutoff artifacts \cite{Litim:2000ci,Litim:2002cf}. \step

Next we employ this reasoning to the data collected in
Tab.~\ref{DE}. A first estimate for the systematic error is obtained
by taking a weighted average over representative entries for each
projection method (standard flow, Wilson-Polchinski flow, background
field flow), disregarding further details of the
approximations. Common to the data points is that the underlying
regulators are, at least partially, optimised \cite{Litim:2000ci}. We
first consider the data points Tab.~\ref{DE}a), b), f), h), k), l), n)
to obtain
\begin{equation}\label{meanall}
\nu=0.631^{+0.018}_{-0.006}\,,\quad
\eta=0.036^{+0.008}_{-0.005}\,,\quad
\omega=0.783^{+0.082}_{-0.127}\,.
\end{equation}
For $\eta$, we only took the data with $\eta\neq 0$ into account. The
mean values \eq{meanall} change by less than $0.1\% (1.5\%)$ for
$\nu,\eta$ $(\omega)$ had we included the data points i), j) and m)
based on some additional input. Hence \eq{meanall} represents an
average with equal weight for the different implementations of the
functional flow. Note that the width of the error bars, roughly a
standard deviation, are set by the least advanced approximations.\step

An improved estimate is obtained by retaining only the most advanced
results in Tab.~\ref{DE}, ie.~k), l) and n), all of which are based on
a similar operator content, supported by a partial cutoff optimisation
\cite{Litim:2000ci}, but differ in the projection technique. We recall
that in k) a standard full fourth-order derivative expansion is used,
together with a polynomial expansion in the fields
\cite{Canet:2003qd}; in n) a mixed approximation is employed retaining
momentum- and field-dependences, but neglecting loop momenta of
certain vertex functions \cite{Benitez:2009xg}; in l) a background
field flow is used within a fourth-order derivative expansion and
without polynomial expansion in the fields, but neglecting higher
order flow terms and subleading fourth-order derivative operators in
the action (this work).  The qualitative differences in the
approximation make sure that the computations project in different
manners onto the Wilson-Fisher fixed point, thereby probing the
systematic error. We find
\begin{equation}\label{mean}
\nu=0.630^{+0.002}_{-0.005}\,,\quad
\eta=0.034^{+0.005}_{-0.003}\,,\quad
\omega=0.823^{+0.043}_{-0.043}\,.
\end{equation}
Note that we omit the data set m) from this estimate to achieve a
conservative error bar and an equal weight between projection
methods. From \eq{meanall} to \eq{mean} the error bars are reduced by
at least a factor of two. The mean values for $\nu$, $\omega$ and
$\eta$ are shifted by $0.2\%$, $5\%$ and $6\%$, respectively. The
shift in the mean values from \eq{meanall} to \eq{mean} is of a
similar size as the estimated error in \eq{mean}. \step

\begin{table}[t]
\begin{center}
\begin{tabular}{clllr}
 \hline \hline
&
$\eta$
&
$\nu$
&
$\omega$
&${}\quad$ref.~/ year${}\quad$
\\
resummed PT
&
0.0335(25) 
&
0.6304(13) 
&
0.799(11) 
&
\cite{Guida:1998bx} \quad{\tiny (1998)}${}\quad$\\
$\eps$-expansion 
&0.0360(50)
&0.6290(25)
&0.814(18)
&
\cite{Guida:1998bx} \quad{\tiny (1998)}${}\quad$\\
world average
&
$0.0364(5)\ \,$
&
$0.6301(4)\ \,$
&
$0.84(4)\ \ \, \,$ 
&
\cite{Pelissetto:2000ek} \quad{\tiny (2000)}${}\quad$\\
Monte Carlo 
&
$0.03627(10)$
&
$0.63002(10)$ 
&
$0.832(6)$
&
\cite{Hasenbusch:2010} \quad{\tiny (2010)}${}\quad$
\\ 
${}\quad$ functional RGs${}\quad$
&
$0.034(5)\ \ \,\,$ 
&
$0.630(5)\ \ \,\,$ 
&
$0.82(4)\ \ \,\,$
&
this work${}\quad$ \\
 \hline \hline
 \end{tabular}
\end{center}
\caption{\label{FRG}Comparison of results from the functional renormalisation
  group  with resummed perturbation theory, Monte-Carlo simulations,
  $\epsilon$-expansion, and a world average.}
\end{table}

In Tab.~\ref{FRG}, the combined functional RG results \eq{mean} are
compared with the $\eps$-expansion, resummed perturbation theory,
Monte Carlo simulations, and a world average of theoretical
predictions.  It shows that the functional RG results agree very well
with results from other methods within systematic errors and on the
level of the mean values. The results are also compatible with recent
experimental results, eg.~$\eta=0.041\pm0.005$ and $\nu=0.632\pm0.002$
\cite{Lytle:2004}, with experimental errors slightly larger than those
from theory (see \cite{Barmatz:2007,Sengers:2009} for
overviews). Expected errors from the functional RG are presently about
an order of magnitude larger than those from eg.~numerical simulations,
and more data and extended approximations are required to further
reduce the systematic uncertainty. In particular, the value for
$\omega$ in \eq{mean} is presently only based on two data
points. Here, it would be useful to know the value from the standard
flow at fourth's order in the derivative expansion to improve the
error estimate in Tab.~\ref{FRG}.  Natural candidates for further data
points are approximations of the Wilson-Polchinski equation beyond
second order in the derivative expansion, or approximations with an
improved access to the full momentum structure of propagator and
vertices.  \\

\noindent{\bf 10. Discussion}\step

The computation of universal scaling exponents is an important
testing ground for methods in quantum field theory and statistical
physics. We have obtained new results for the indices $\nu$, $\omega$,
$\omega_5$ and $\eta$ of the 3d Ising universality class
 using functional renormalisation group
methods within a background field formulation. Our analysis
complements earlier studies without background fields. Our findings to
second \eq{eta2nd} and fourth order \eq{4th}, \eq{opt4} in the
derivative expansion agree very well with other theoretical studies. A
fast numerical convergence of the derivative expansion is found,
confirming similar observations to lower order in the expansion. The
indices also display the correct cross-dependences. This non-trivial result 
lends further support to the underlying approximations. \step

We have also studied the cross-correlations of exponents, and their sensitivity 
to tiny variations of the dimensionality. The latter correlates with the expected
error of exponents within the derivative expansion. As a result \eq{A}, the 
index $\nu$ shows a weak, the subleading index $\omega$ a moderate, 
and the anomalous dimension a strong dependence on dimensional 
variations. We conclude that the achievable precision in these observables 
follows the same pattern, as confirmed by the data \eq{mean}. \step

The flexibility of the functional renormalisation group allows for
different projections onto the Wilson-Fisher fixed point.  We have
exploited this freedom to estimate the systematic uncertainty of
scaling exponents using all available data. 
The resulting mean values and error estimates
\eq{mean} agree very well with results from resummed perturbation
theory and lattice simulations.  More work and further data points are
required to reduce the error bars, which are similar to those from
experiment, but larger than those from recent numerical
simulations. Natural candidates for further data points are 
eg.~Wilson-Polchinski flows to fourth 
order in the derivative expansion, and  approximation schemes with an 
improved access to the momentum structure of propagators and vertices.
\step

In addition, we have analysed the convergence of the derivative
expansion, comparing data from standard flows, background
field flows, and the Wilson-Polchinski flow. Background field flows lead
systematically to smaller values for $\eta$,
and the derivative expansion converges very fast. Standard flows 
provide narrower bounds on exponents, while the derivative expansion shows 
a slightly slower rate of convergence. 
For the Wilson-Polchinski flow, structural arguments suggest that approximations beyond 
the leading order are more sensitive to the cutoff. Still, good results are available
to second order, provided $\eta$ is matched. 
It will thus be interesting to extend these studies 
beyond the Ising universality class.  \\[2ex]

\noindent{\bf Acknowledgements}\step

DL thanks Anna Hasenfratz and Boris Svistunov for discussions, Martin
Hasenbusch for e-mail correspondence, and the Aspen Center for Physics
for hospitality.


\begin{thebibliography}{99}


\bibitem{ZinnJustin}
J.~Zinn-Justin, {\it Quantum field theory and 
critical phenomena}, Oxford (Clarendon, 1989).

\bibitem{Pelissetto:2000ek}
  A.~Pelissetto and E.~Vicari,
  {\it Critical phenomena and renormalization-group theory},
  Phys.\ Rept.\  {\bf 368} (2002) 549
  [cond-mat/0012164].

\bibitem{Polchinski:1983gv}
  J.~Polchinski,
  {\it Renormalization and Effective Lagrangians},
  Nucl.\ Phys.\  {\bf B231 } (1984)  269-295.

\bibitem{Wetterich:1992yh}
  C.~Wetterich,
  {\it Exact evolution equation for the effective potential},  
Phys.\ Lett.\  B {\bf 301} (1993) 90.
  
\bibitem{Wilson:1973jj}
  K.~G.~Wilson, J.~B.~Kogut,
  {\it The Renormalization group and the epsilon expansion},
  Phys.\ Rept.\  {\bf 12 } (1974)  75-200.
  
\bibitem{Litim:2000ci}
D.~F.~Litim,
{\it Optimisation of the exact renormalisation group},
Phys.\ Lett.\  B {\bf 486} (2000) 92, [hep-th/0005245].

\bibitem{Litim:2001up}
  D.~F.~Litim,
{\it Optimised renormalisation group flows},
  Phys.\ Rev.\  D {\bf 64} (2001) 105007
  [hep-th/0103195].

\bibitem{Litim:2001fd}
  D.~F.~Litim,
  {\it Mind the gap},
  Int.\ J.\ Mod.\ Phys.\  A {\bf 16} (2001) 2081
  [hep-th/0104221].

\bibitem{Litim:2002cf}
D.~F.~Litim,
{\it Critical exponents from optimised renormalisation group flows},
Nucl.\ Phys.\ B {\bf 631} (2002) 128, [hep-th/0203006].

\bibitem{Pawlowski:2005xe}
  J.~M.~Pawlowski,
  {\it Aspects of the functional renormalisation group},
  Annals Phys.\  {\bf 322 } (2007)  2831-2915.
  [hep-th/0512261].

  
  \bibitem{Bagnuls:2000ae}
  C.~Bagnuls, C.~Bervillier,
  {\it Exact renormalization group equations. An Introductory review},
  Phys.\ Rept.\  {\bf 348 } (2001)  91.
  [hep-th/0002034].

  
\bibitem{Berges:2000ew}
  J.~Berges, N.~Tetradis, C.~Wetterich,
  {\it Nonperturbative renormalization flow in quantum 
field theory and statistical physics},
  Phys.\ Rept.\  {\bf 363 } (2002)  223-386.
  [hep-ph/0005122].

\bibitem{Polonyi:2001se}
  J.~Polonyi,
  {\it Lectures on the functional renormalization group method},
  Central Eur.\ J.\ Phys.\  {\bf 1} (2003) 1
  [hep-th/0110026].
  
\bibitem{Litim:1998nf}
  D.~F.~Litim, J.~M.~Pawlowski,
  {\it On gauge invariant Wilsonian flows}, 
in: The Exact Renormalization Group, Eds. Krasnitz et al,
World Sci (1999) 168    [hep-th/9901063].

\bibitem{Litim:2005us}
  D.~F.~Litim,
  {\it Universality and the renormalisation group},
  JHEP {\bf 0507} (2005) 005
  [hep-th/0503096].

\bibitem{Litim:2001ky}
  D.~F.~Litim and J.~M.~Pawlowski,
  {\it Perturbation theory and renormalisation group equations},
  Phys.\ Rev.\  D {\bf 65} (2002) 081701
  [hep-th/0111191].

\bibitem{Litim:2002xm}
  D.~F.~Litim and J.~M.~Pawlowski,
  {\it Completeness and consistency of renormalisation group flows},
  Phys.\ Rev.\  D {\bf 66} (2002) 025030
  [hep-th/0202188].


\bibitem{Litim:2002hj}
  D.~F.~Litim and J.~M.~Pawlowski,
  {\it Wilsonian flows and background fields},
  Phys.\ Lett.\  B {\bf 546} (2002) 279
  [hep-th/0208216].
  
\bibitem{Litim:2006nn}
  D.~F.~Litim, J.~M.~Pawlowski and L.~Vergara,
{\it Convexity of the effective action from functional flows},
[hep-th/0602140].

\bibitem{Liao:1996fp}
S.~B.~Liao,
{\it On connection between momentum cutoff and the
proper time regularizations},
Phys.\ Rev.\  {\bf D53} (1996) 2020,
[hep-th/9501124].


\bibitem{Bohr:2000gp}
  O.~Bohr, B.~J.~Schaefer, J.~Wambach,
  {\it Renormalization group flow equations and 
the phase transition in O(N) models},
  Int.\ J.\ Mod.\ Phys.\  {\bf A16 } (2001)  3823-3852.
  [hep-ph/0007098].

\bibitem{Bonanno:2000yp}
  A.~Bonanno and D.~Zappal\`a,
  {\it Towards an accurate determination of the critical 
exponents with the  renormalization group flow equations},
  Phys.\ Lett.\  B {\bf 504} (2001) 181,
  [hep-th/0010095].

\bibitem{Mazza:2001bp}
M.~Mazza and D.~Zappal\`a,
 {\it Proper time regulator and renormalization group flow},
Phys.\ Rev.\ D {\bf 64} (2001) 105013, [hep-th/0106230].

\bibitem{Litim:2001hk}
D.~F.~Litim and J.~M.~Pawlowski,
{\it Predictive power of renormalisation group flows: A comparison},
Phys.\ Lett.\ B {\bf 516} (2001) 197, [hep-th/0107020].

\bibitem{Meurice:2007zg}
  Y.~Meurice,
  {\it Nonlinear Aspects of the Renormalization Group Flows 
of Dyson's Hierarchical Model},
  J.\ Phys.\ A {\bf A40 } (2007)  R39.
  [hep-th/0701191].

\bibitem{Litim:2007jb}
  D.~F.~Litim,
  {\it Towards functional flows for hierarchical models},
  Phys.\ Rev.\  D {\bf 76} (2007) 105001
  [0704.1514 [hep-th]].

\bibitem{Morris:1994ie}
  T.~R.~Morris,
  {\it Derivative expansion of the exact renormalization group},
  Phys.\ Lett.\  B {\bf 329}, 241 (1994)
  [hep-ph/9403340].

\bibitem{Litim:2001dt}
  D.~F.~Litim,
  {\it  Derivative expansion and renormalisation group flows},
  JHEP {\bf 0111} (2001) 059
  [hep-th/0111159].

\bibitem{Blaizot:2005xy}
  J.~-P.~Blaizot, R.~Mendez Galain, N.~Wschebor,
  {\it A New method to solve the non perturbative 
renormalization group equations},
  Phys.\ Lett.\  {\bf B632 } (2006)  571-578.
  [hep-th/0503103].

\bibitem{Litim:1998yn}
  D.~F.~Litim,
  {\it Wilsonian flow equation and thermal field theory},
  hep-ph/9811272.

\bibitem{Bergerhoff:1995zq}
  B.~Bergerhoff, F.~Freire, D.~Litim, S.~Lola and C.~Wetterich,
  {\it Phase diagram of superconductors}, 
  Phys.\ Rev.\  B {\bf 53} (1996) 5734
  [hep-ph/9503334].
  
\bibitem{Gies:2002af}
  H.~Gies,
  {\it Running coupling in Yang-Mills theory: A flow equation study},  
  Phys.\ Rev.\  D {\bf 66}, 025006 (2002)
  [hep-th/0202207].

\bibitem{Pawlowski:2003hq}
  J.~M.~Pawlowski, D.~F.~Litim, S.~Nedelko, and L.~v.~Smekal,
  {\it Infrared behavior and fixed points in Landau gauge QCD},
  Phys.\ Rev.\ Lett.\  {\bf 93 } (2004)  152002.
  [hep-th/0312324].

\bibitem{Litim:2008tt}
  D.~F.~Litim,
  {\it Fixed Points of Quantum Gravity and the 
Renormalisation Group},  [0810.3675 [hep-th]].

\bibitem{Reuter:1993kw}
  M.~Reuter and C.~Wetterich,
  {\it Effective average action for gauge theories and exact evolution
  equations},
  Nucl.\ Phys.\  B {\bf 417} (1994) 181.
  
\bibitem{Freire:2000bq}
  F.~Freire, D.~F.~Litim and J.~M.~Pawlowski,
  {\it Gauge invariance and background field formalism in the exact
  renormalisation group},
  Phys.\ Lett.\  B {\bf 495} (2000) 256
  [hep-th/0009110].
  
\bibitem{Litim:2002ce}
  D.~F.~Litim and J.~M.~Pawlowski,
  {\it Renormalisation group flows for gauge theories in axial gauges},
  JHEP {\bf 0209} (2002) 049
  [hep-th/0203005].

\bibitem{Litim:1996nw}
  D.~F.~Litim,
  {\it Scheme independence at first order phase transitions and the
  renormalisation group},
  Phys.\ Lett.\  B {\bf 393} (1997) 103
  [hep-th/9609040].

\bibitem{Freire:2000sx}
  F.~Freire and D.~F.~Litim,
  {\it Charge cross-over at the U(1)-Higgs phase transition},
  Phys.\ Rev.\  D {\bf 64} (2001) 045014
  [hep-ph/0002153].


  
\bibitem{Zappala:2001nv}
  D.~Zappal\`a,
  {\it Improving the Renormalization Group approach to the quantum-mechanical
  double well potential},
  Phys.\ Lett.\  A {\bf 290}, 35 (2001)
  [quant-ph/0108019].

\bibitem{Bonanno:2004pq}
  A.~Bonanno and G.~Lacagnina,
  {\it Spontaneous symmetry breaking and proper-time flow equations},
  Nucl.\ Phys.\  B {\bf 693}, 36 (2004)
  [hep-th/0403176].

\bibitem{Consoli:2006ji}
  M.~Consoli and D.~Zappal\`a,
  {\it Renormalization-group flow for the field strength in scalar
  self-interacting theories},
  Phys.\ Lett.\  B {\bf 641}  (2006) 368
  [hep-th/0606010].

\bibitem{Castorina:2003kq}
  P.~Castorina, M.~Mazza and D.~Zappal\`a,
  {\it Renormalization group analysis of the three-dimensional
  Gross-Neveu  model at finite temperature and density},
  Phys.\ Lett.\  B {\bf 567}  (2003) 31
  [hep-th/0305162].

\bibitem{Bonanno:2004sy}
  A.~Bonanno and M.~Reuter,
  {\it Proper time flow equation for gravity},
  JHEP {\bf 0502} (2005) 035
  [hep-th/0410191].

\bibitem{Zappala:2002nx}
  D.~Zappal\`a,
  {\it Perturbative and non-perturbative aspects of the proper
   time  renormalization group}",
   Phys.\ Rev.\  D {\bf 66} (2002) 105020
   [hep-th/0202167].

\bibitem{Ballhausen:2003bu}
  H.~Ballhausen,
  {\it The effective average action beyond first order},
  [hep-th/0303070].

\bibitem{Canet:2002gs}
  L.~Canet, B.~Delamotte, D.~Mouhanna and J.~Vidal,
  {\it Optimization of the derivative expansion in the nonperturbative
  renormalization group},
  Phys.\ Rev.\  D {\bf 67} (2003) 065004
  [hep-th/0211055].

\bibitem{Canet:2003qd}
  L.~Canet, B.~Delamotte, D.~Mouhanna and J.~Vidal,
  {\it Nonperturbative renormalization group approach to the Ising model: a
  derivative expansion at order $\partial^4$},
  Phys.\ Rev.\  B {\bf 68} (2003) 064421
  [hep-th/0302227].
  
\bibitem{Litim:2003kf}
  D.~F.~Litim and L.~Vergara,
 {\it Subleading critical exponents from the renormalisation group},
  Phys.\ Lett.\  B {\bf 581} (2004) 263,
  [hep-th/0310101].

\bibitem{Hasenbusch:2010}
M.~Hasenbusch, {\it A Finite Size Scaling Study of 
Lattice Models in the 3D Ising Universality Class}, [1004.4486].

\bibitem{Ballhausen:2003gx}
  H.~Ballhausen, J.~Berges and C.~Wetterich,
  {\it Critical phenomena in continuous dimension},
  Phys.\ Lett.\  B {\bf 582}, 144 (2004)
  [hep-th/0310213].

\bibitem{Guida:1998bx}
  R.~Guida and J.~Zinn-Justin,
 {\it Critical Exponents of the N-vector model},
  J.\ Phys.\ A  {\bf 31} (1998) 8103,
  [cond-mat/9803240].
  
  \bibitem{Hasenfratz:1985dm}
  A.~Hasenfratz and P.~Hasenfratz,
  {\it Renormalization Group Study Of Scalar Field Theories},
  Nucl.\ Phys.\  B {\bf 270} (1986) 687
  [Helv.\ Phys.\ Acta {\bf 59} (1986) 833].
  
\bibitem{Bervillier:2007rc}
  C.~Bervillier, A.~Juttner and D.~F.~Litim,
  {\it High-accuracy scaling exponents in the local potential approximation},
  Nucl.\ Phys.\  B {\bf 783} (2007) 213,
  [hep-th/0701172].

\bibitem{Seide:1998ir}
  S.~Seide, C.~Wetterich,
  {\it Equation of state near the endpoint of the critical line},
  Nucl.\ Phys.\  {\bf B562 } (1999)  524-546.
  [cond-mat/9806372].
  
\bibitem{VonGersdorff:2000kp}
  G.~Von Gersdorff, C.~Wetterich,
  {\it Nonperturbative renormalization flow and 
essential scaling for the Kosterlitz-Thouless transition},
  Phys.\ Rev.\  {\bf B64 } (2001)  054513.
  [hep-th/0008114].

\bibitem{Bervillier:2005za}
  C.~Bervillier,
  {\it Wilson-Polchinski exact renormalization group 
equation for O(N) systems:   Leading and next-to-leading 
orders in the derivative expansion},
  J.\ Phys.\ Condens.\ Matter {\bf 17} (2005) S1929
  [hep-th/0501087].

\bibitem{Ball:1994ji}
  R.~D.~Ball, P.~E.~Haagensen, J.~I.~Latorre and E.~Moreno,
  {\it Scheme Independence And The Exact Renormalization Group},
  Phys.\ Lett.\  B {\bf 347} (1995) 80
  [hep-th/9411122].

\bibitem{Comellas:1997ep}
  J.~Comellas,
  {\it Polchinski equation, reparameterization invariance 
and the derivative  expansion},
  Nucl.\ Phys.\  B {\bf 509} (1998) 662
  [hep-th/9705129].
  
\bibitem{Benitez:2009xg}
  F.~Benitez, J.~P.~Blaizot, H.~Chate, B.~Delamotte, 
R.~Mendez-Galain and N.~Wschebor,  {\it Solutions of 
renormalization group flow equations with full momentum  dependence},
  Phys.\ Rev.\  E {\bf 80} (2009) 030103   
[0901.0128 [cond-mat.stat-mech]].

\bibitem{Stevenson:1981vj}
  P.~M.~Stevenson,
  {\it Optimized Perturbation Theory},
  Phys.\ Rev.\  {\bf D23 } (1981)  2916.
  
\bibitem{Liao:1999sh}
  S.~-B.~Liao, J.~Polonyi, M.~Strickland,
  {\it Optimization of renormalization group flow},
  Nucl.\ Phys.\  {\bf B567 } (2000)  493-514.
  [hep-th/9905206].

\bibitem{Lytle:2004} A. Lytle and D. T. Jacobs,  
{\it Turbidity determination of the critical 
exponent $\nu$ in the liquid Ð liquid 
mixture methanol and cyclohexane}, 
J. Chem. Phys. 120 (2004) 5709.

\bibitem{Barmatz:2007}  
M. Barmatz, I. Hahn, J. A. Lipa, and R. V. Duncan, 
{\it Critical phenomena in microgravity: 
Past, present, and future}, Rev. Mod. Phys. 79 (2007) 1.

\bibitem{Sengers:2009} J. V. Sengers and J. G. Shanks, 
{\it Experimental Critical-Exponent Values for Fluids} 
J. Stat. Phys. 137 (2009) 857  
  
    \end{thebibliography}
\end{document}